# A FUTURE MOBILE PACKET CORE NETWORK BASED ON IP-IN-IP PROTOCOL

Mohammad Al Shinwan[1] and Kim Chul-Soo[2]

[1]Faculty of Computer Science and Informatics, department of Mobile Computing, Amman Arab University, Amman, Jordan.
[2]Department of Computer Engineering, Inje University, Gimhae, Republic of Korea.

## ABSTRACT

*The current Evolved Packet Core (EPC) 4th generation (4G) mobile network architecture features complicated control plane protocols and requires expensive equipment. Data delivery in the mobile packet core is performed based on a centralized mobility anchor between eNode B (eNB) elements and the network gateways. The mobility anchor is performed based on General Packet Radio Service tunnelling protocol (GTP), which has numerous drawbacks, including high tunnelling overhead and suboptimal routing between mobile devices on the same network. To address these challenges, here we describe new mobile core architecture for future mobile networks. The proposed scheme is based on IP encapsulated within IP (IP-in-IP) for mobility management and data delivery. In this scheme, the core network functions via layer 3 switching (L3S), and data delivery is implemented based on IP-in-IP routing, thus eliminating the GTP tunnelling protocol. For handover between eNB elements located near to one another, we propose the creation of a tunnel that maintains data delivery to mobile devices until the new eNB element updates the route with the gateway, which prevents data packet loss during handover. For this, we propose Generic Routing Encapsulation (GRE) tunnelling protocol. We describe the results of numerical analyses and simulation results showing that the proposed network core architecture provides superior performance compared with the current 4G architecture in terms of handover delay, tunnelling overhead and total transmission delay.*

## 1. INTRODUCTION

In recent years, the marked increase in the use of smart phones and other mobile devices has led to huge growth in wireless mobile communication data traffic. This trend appears likely to continue, and Cisco forecasts that the volume of mobile data traffic will increase eight-fold between 2015 and 2020 [1]. This growth in mobile data traffic places increasing demands on wireless communication systems, and represents a major challenge for cellular providers in terms of upgrading their core networks to accommodate future network requirements and keeping up with increasing customer demand.

One of the greatest challenges for future mobile communication networks is how to design and build 5th generation (5G) mobile networks. The need for new network architecture is essential to support growth in demand for broadband services of various kinds delivered over the networks, and to support the Internet of Things (IoT) services and applications [2].

Many approaches have been proposed to address the growth in data traffic on mobile networks, including device-to-device communication and radio resource management. However, these efforts have focused mainly on increasing the capacity of wireless radio links. The future mobile network consists of two main parts: a radio link and a non-radio mobile core network. Effective design of both the radio link and the mobile core is required to meet the requirements of the future mobile network [3].

The current 4th generation (4G) core network termed the Evolved Packet Core (EPC) is based on the General Packet Radio Service tunnelling protocol (GTP) [4]. With EPC, eNodeB (eNB) elements establish GTP tunnels with serving gateways (SGWs) and Packet Data Network (PDN) gateways (PGWs) to create centralized mobility anchors for data packet forwarding. However, the 4G network has a number of limitations. First, there are load balance and latency issues. Growth in data traffic requires a reduction in the transmission and connection delays. Simplifying the mobile core and reducing the number of identifications can make mobile core networks simpler and more efficient, and hence more cost-effective. The second problem is suboptimal routing. With current 4G networks, the uplink and downlink data packets are routed via mobility anchors, which often results in the suboptimal paths. For example, a data instead of taking the shortest path, the packet is routed via a PGW and an SGW. The Third is the GTP tunnelling protocol overhead. GTP protocol adds three headers to the data payload totalling 36 bytes; i.e., GTP,

User Datagram Protocol (UDP) and IP. In addition, the use of GTP protocol and mobility anchors means that the full functionality of packet switching cannot be exploited, the result being that circuit switching is favoured over packet switching. Fourth is the required capital expenditure. The EPC network is simply not cost-effective, due to the huge number of routers that are required to support the core network.

## 2. LITERATURE REVIEW

A variety of schemes have been proposed to overcome these issues. The Distributed Mobility Management (DMM) scheme proposed by the Internet Engineering Task Force [5] provides mobility solutions with localized mobility anchors that are distributed within the network, in combination with centralized anchors, where the system is arranged in a hierarchical model. This was proposed to optimize routing for local data traffic, and reduces delays due to the shorter distances to local servers [6]. A mobility data offloading approach using femtocells has been proposed to enhance the 4G network [7]. In this scheme, data traffic is forward to the mobile device without using the mobile core network, which reduces the volume of internet traffic and unwanted data flow into the mobile core network.

Another approach to distributing mobility in 4G networks is the Ultra Flat Architecture (UFA). The key element of UFA is to decrease the number of the network nodes to one (i.e., a single base station), based on the distribution of user and control plane roles in the node. UFA provides improved performance and seamless handover [8]. In [9] and [10], the authors proposed a new mobile network architecture for 5G networks termed 5G-TPC. This architecture was based on the Transparent Interconnection of Lots of Links (TRILL) protocol, and was designed to use link layer routing bridges rather than GTP protocol.

## 3. METHODOLOGY

Most existing (including previously proposed) architectures suffer from problems associated with mobility anchoring and GTP overhead. Moreover, from the perspective of capital expenditure, most of the proposed mobile network architectures are not cost-effective. This is because the core network is based on routers. Here we propose to use layer 3 switching (L3S), which has significant advantages both in terms of cost and performance, provides high performance and can handle several different routing protocols, including Open Shortest Path First (OSPF), Intermediate System to Intermediate System (IS-IS) and Routing Information Protocol (RIP) [11]. Layer 3 switches are cheaper than routers, and can exploit the advantages of layers 2 and 3.

In this paper, we also propose a new architecture for the future mobile core network, termed ICNA (IP-in-IP Core Network Architecture). The proposed scheme is based on IP encapsulated within IP (IP-in-IP) [12], and aims to eliminate the GTP tunnelling overhead and simplify the mobile packet core between the eNB and PGW. The proposed approach also eliminates the legacy centralized mobility anchor through the use of distributed data traffic and data routing within the mobile network using layer 3 switches, and exploits the full functionality of packet switching in the mobile core network. The approach entails the use of two IP addresses: an inner IP address for the user equipment (UE) identifier, and an outer IP address for the eNB or PGW identifier.

For handover between eNBs located near to one another, we propose the use of a tunnel to maintain data flow to mobile devices until the new eNB updates the route with the gateway. In this manner, we prevent data packet loss during handover. For this, we propose Generic Routing Encapsulation (GRE) tunnelling protocol.

The remainder of the paper is organized as follows. Section 2 reviews the 4G mobile core network architecture. Section 3 describes the proposed ICNA scheme, and gives an overview of the handover process. In Section 4, we compare the 4G network with the proposed ICNA network via a numerical analysis, and give a performance evaluation. Section 5 concludes the paper.

## 4. MOBILITY MANAGEMENT IN 4G NETWORKS

System Architecture Evolution (SAE) [13] is the non-radio core architecture of Long Term Evolution (LTE) networks developed by 3GPP. It is an evolution of the General Packet Radio Service (GPRS) network, and aims to support low latency, high throughput and mobility between multi-heterogeneous networks. The most important component of the SAE architecture is the EPC. EPC networks are formed of several functional entities, namely PGWs, SGWs, Mobility Management Entities (MMEs), Home Subscriber Servers (HSSs), and Policy and Charging Rules Function (PCRF) servers. PGWs provide mobile users with access to a PDN by allocating IP addresses, and also provide IP routing and forwarding; SGWs act as local mobility anchors for inter-eNB handover; MMEs provide several functions, including mobility management and handover management; HSSs provide user profiles and authentication data; and PCRF controls the charging rules and quality of service [19].

In an EPC network, data paths are established between eNBs and PGWs via SGWs, and use GTP for tunnelling. PGWs and SGWs behave as centralized mobility anchors for data packets; consequently, all data traffic is forward through a centralized anchor SGW and PGW.

As shown in Figure (1), the EPC architecture is composed of several interfaces for data forwarding, including X2, S1-MME, S11, S5 and S1-U. The X2 interface protocol supports UE mobility by creating a GTP tunnel between eNBs. S1-MME provides the initial UE context to MMEs, and is also responsible for establishing and controlling the GTP tunnel between MMEs and SGWs. S5 provides GTP tunnel functionality between SGWs and PGWs. The S1-U interface provides GTP tunnel function between eNBs and SGWs.

Figure (2) shows the protocol stack for data delivery based on GTP. When the UE sends a data packet, the eNB adds IP/UDP/GTP headers. GTP protocol is used for data that flows between eNBs, SGWs and PGWs. When a PGW receives a data packet, it removes all GTP headers, and then forwards the packet to the Internet host.

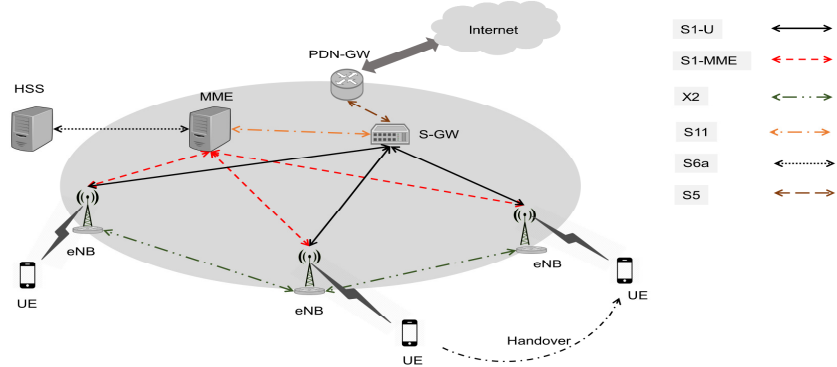

Figure 1. The Evolved Packet Core (EPC) model for 4th generation (4G) mobile networks.

The data delivery process in the 4G network is shown in Figures (3) and (4). The UE sends a packet to a PGW via an eNB and an SGW. The PGW receives the data packet and determines the location of the destination from its database. PGW determines whether the destination is within the same mobile network or is outside of it, and then forwards the data packet to the destination.

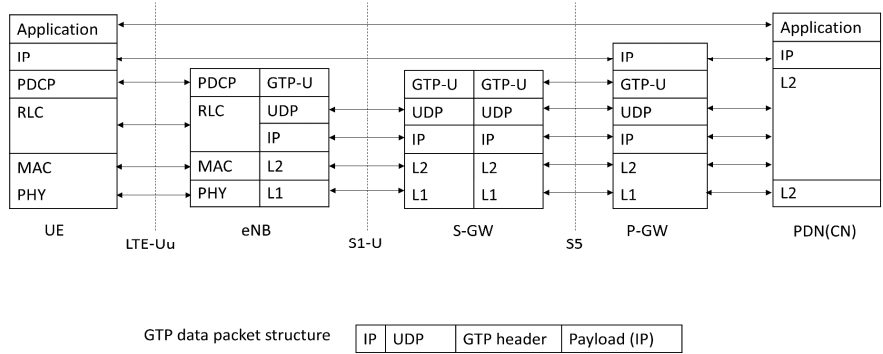

Figure 2. The protocol stack for data delivery in a 4G network.

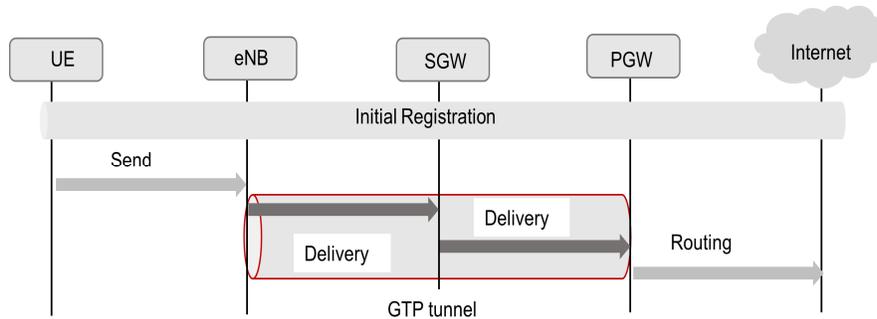

Figure 3. Data delivery from a UE to the Internet.

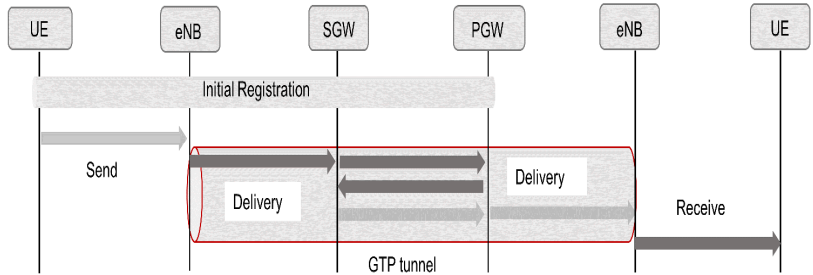

Figure 4. Data delivery from a user equipment (UE) to another UE.

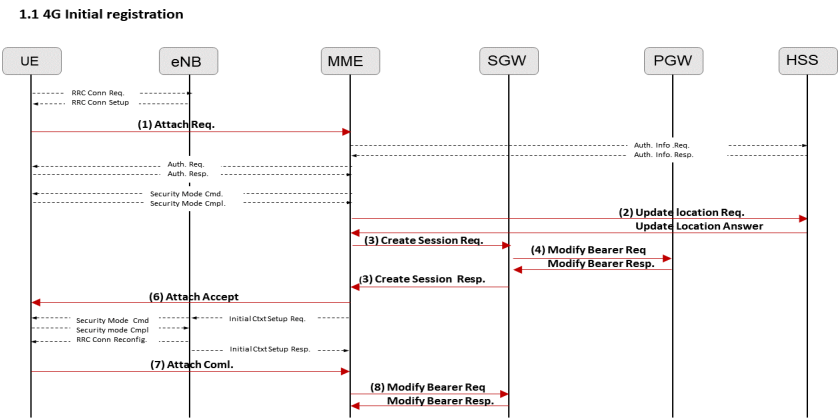

Figure 5. The initial attach procedure in a 4G network.

Figure (5) shows the initial attach registration procedure for a 4G network. When the UE establishes radio link synchronization with an eNB, the UE creates a connection for data delivery via an Attach Request message sent to the eNB. The eNB then forwards the attach request to an MME, which sends an Update Location Request to an HSS. The HSS responds via an Update Location Answer, and then the MME performs the required security related-operations with UE. The MME sends a Create Session Request to an SGW to create a transmission path. The SGW now sends a Modify Bearer Request to a PGW, which response by sending a Modify Bearer Response message. The SGW then sends a Create Session Response to the MME, and the MME sends an Attach Accept message to the UE. The MME now performs an Initial Context Setup with the eNB, and the eNB sends an Initial Context Setup Response message to the MME. The UE then sends an Attach Complete message to the MME, which sends a Modify Bearer Request to the SGW, which response by sending a Modify Bearer Response message to the MME.

There are two types of handover process in a 4G network: X2 handover with SGW relocation, and S1 handover with SGW relocation. Figure (6) shows a X2 handover with SGW relocation. When the UE moves to another eNB region, the source eNB sends a Handover Request to the target eNB, which response with a Handover Acknowledgment (step 1). The target eNB then sends a Path Switch Request to the MME (step 2), which sends a Create Session Request to the SGW (step 3). The SGW exchanges modify bearer messages with the PGW via a Modify Bearer Request and a Modify Bearer Response (step4), and then sends a Create Session Response to the MME (step 5), which sends a Path Switch Response to the target eNB (step 6). Finally, the MME sends a Release Resources message to the source eNB (step 7).

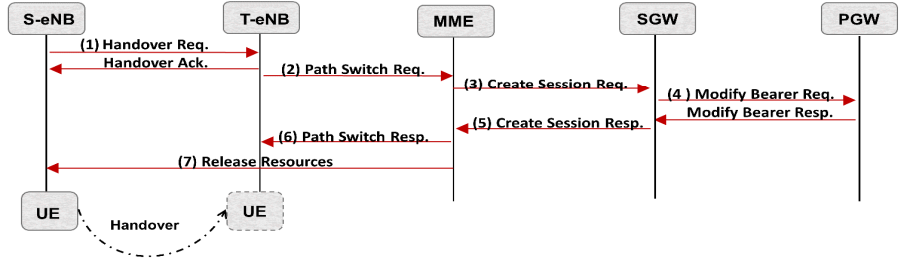

Figure 6. X2 handover with serving gateway (SGW) relocation.

Figure (7) shows the S1 handover with SGW relocation. When the UE moves to a new domain, the source eNB sends a Handover required message to the MME (step 1). The MME then sends a Handover Request to target eNB (step 2), which response with a Handover Acknowledgment (step 3). The MME then sends a Handover Command to the source eNB (step 4), which sends a Handover Notify message to the MME (step 5). The MME sends a Modify Bearer Request to the target SGW (step 6), which exchanges modify bearer messages with the PGW via a Modify Bearer Request and a Modify Bearer Response (step 7). The target SGW sends a Modify Bearer Response to the MME (step 8), which then sends a Release Resources message to the source eNB (step 9).

## 5. PROPOSED NETWORK ARCHITECTURE

### 5.1. Network Model

Figure (8) shows an overview of the proposed ICNA network. Each base station (BS) functions based on layer 3 routing, as does the Cellular Gateway (CGW) for Internet hosts. Layer 3 switches are used for packet delivery in the mobile packet core. For UE identification, an inner IP address (i.e., the original IP address) will be located by the User Control Entity (UCE). The outer IP address of the BS is used as a location reference for the UE. UCE with HSS is used for UE registration and to obtain subscription information on the UE. UCE with HSS is also used to register the UCE ID and indicate in which UCE the UE is located. The CGW is located between the mobile packet core and the Internet.

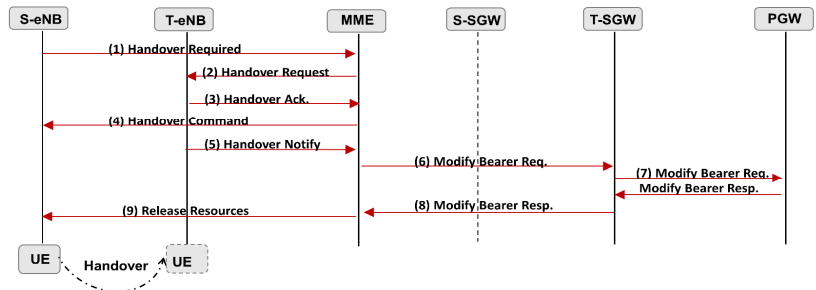

Figure 7. S1 handover with SGW relocation.

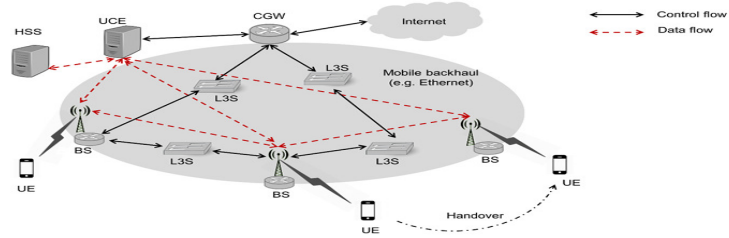

Figure 8. The proposed IP encapsulated within IP (IP-in-IP) packet core for future mobile networks.

The link state protocols OSPF and IS-IS are used for routing in the mobile backhaul, as specified by Cisco for L3S [11]. The full functionality of packet switching networks can therefore be exploited for the link state protocol of the communication system. The GRE tunnelling is used for connection between BSs, and provides support for handover between BSs, preventing packet loss during handover.

## 5.2. Comparison of 4G and ICNA

Table 1 lists the main characteristics of the existing 4G network architecture and the proposed network architecture. With 4G, data delivery uses the GTP protocol, whereas ICNA uses IP-in-IP based on L3S. With 4G, the tunnel endpoint identifier (TEID) of GTP is used as a locator, whereas the outer IP address is used as a locator with ICNA. With 4G, GTP/UDP/IP headers are used for data packet encapsulation for creating GTP tunnel, whereas ICNA uses IP-in-IP encapsulation for data delivery. For UE handover, with 4G the GTP tunnels are re-established between PGWs and eNBs via an SGW.

However, with ICNA, when the UE moves to a new location, information on the UE is updated by the UCE, including the new outer IP address (i.e., new BS IP address). With 4G networks, GTP tunnelling via SGW and PGW determines the data path; thus, data packets may follow suboptimal paths, including the data path between UEs in the same mobile network, as well as between the PGW and UE. With ICNA, however, the data path will be optimal because data packets are forwarded directly between switches using routing protocols.

## 5.3. IP-in-IP Protocol Stack for Data Delivery

Figure (9) shows IP-in-IP protocol stacks for data delivery with communication between a UE and an Internet host. Here IP-in-IP is shown between a BS and a CGW. The radio interface between the BS and UE uses RLC and PDCP protocols. BS encapsulates the IP address of the UE as an inner IP address with an IP-in-IP packet using the outer IP address (i.e., the IP address of the CGW). The switches forward data packets in the mobile core network from the BS to the Internet access point (i.e., CGW). As shown in Figure (10), the outer IP address is used for delivery by the mobile packet core.

Table 1. Comparison of 4G and ICNA

| Item | 4G | ICNA |
|---|---|---|
| Data delivery protocol | GTP | IP-in-IP |
| Locator | TEID (GTP) | Outer IP |
| Encapsulation | IP/UDP/GTP | Outer IP and Inner IP |
| IP address allocation | PDN-GW | UCE |
| Handover update | GTP tunnel re-established | Outer IP update to MME |
| Optimal Path | No | Yes |

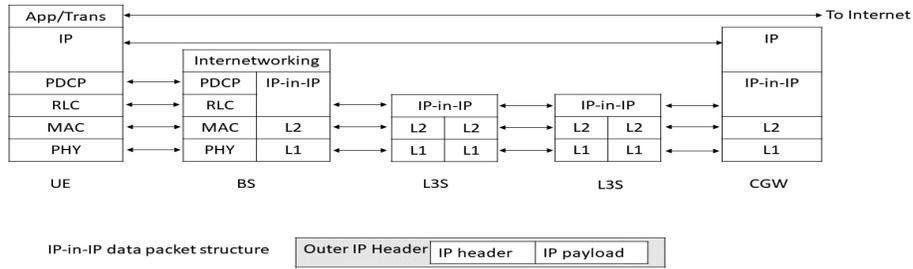

Figure 9. The IP-in-IP protocol stack used for data delivery to the Internet.

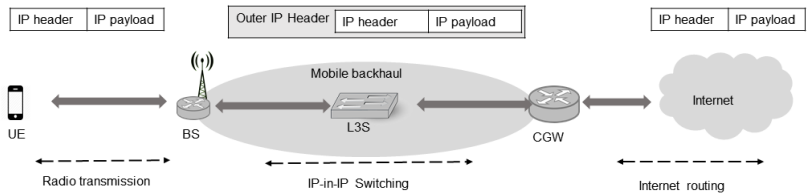

Figure 10. IP-in-IP switching for data delivery to from user equipment (UE) to an Internet host.

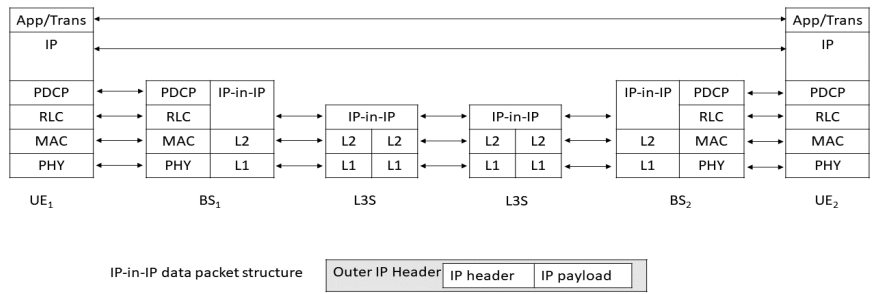

Figure 11. The IP-in-IP protocol stack used for data delivery to UE.

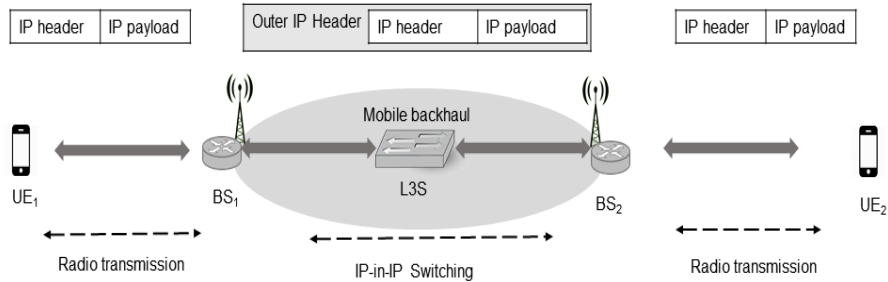

Figure 12. The IP-in-IP switching used for data delivery to from user equipment to UE.

Figure (11) shows the protocol stack and IP-in-IP switching used for communication between two UEs in the same mobile network (UE1 and UE2). Here UE1 is connected to BS1 and UE2 is connected to BS2, and data access between the UEs and the BSs is via RLC and PDCP protocols. When BS1 receives a Packet Send Request from UE1 to obtain the address of BS2, BS1 will encapsulate the original packet with the IP address of BS2. The packet is then forwarded to BS2 via switches, and BS2 de-encapsulates the packet, extracts the inner IP address, and sends it to UE2. Figure (12) shows how the inner and outer IP addresses are used for packet delivery between UE1 and UE2.

### 5.4. Initial Attach Procedure

Figure (13) shows the initial attach registration procedure used with ICNA. When the UE establishes radio link synchronization with the BS, the UE sends an Attach Request message to the BS (step 1). The BS then sends the Attach Request to the UCE, which sends an Authentication Information to the HSS, as well as a Network Attach Storage (NAS) request. Once these authentications and NAS security procedures are accomplished, the UCE sends an Update Location Request to the HSS, which indicates in which UCE the UE is located. The HSS responds by sending an Update Location Answer (step 2).

In step (3), the UCE sends a request to the CGW to allocate a gateway address for the UE via exchange of Gateway Allocation Request and Gateway Allocation Response messages. An IP address is then allocated to the UE by the BS via exchange of an IP Allocation Request and an IP Allocation Response (step 4). Following the allocation of the IP address and establishment of a gateway, the UCE responds to the BS with an Attach Accept message, which contains the IP addresses of the BS and the IP (i.e., the outer and inner IP addresses). The BS then sends an Attach Accept message to the UE (step 5), and the UE sends an Attach Complete message to the BS (step 6).

### 5.5. Data Delivery Procedures

Once the initial attach process has been completed, the UE can send and receive data packets. Data delivery in ICNA is categorized as either mobile host to Internet host, mobile host to mobile host, or Internet host to mobile host.

#### 5.5.1. Mobile Host to Internet Host

Figure (14) shows the data delivery procedure for data transfer from a mobile host to an Internet host. First, the UE sends a data packet to the BS, which identifies whether or not the destination belongs to the same mobile network by checking the IP address space. Where the destination IP address is outside of the mobile network, the BS encapsulates the data packet with the IP address of the CGW, and then forwards it to the CGW, which de-encapsulates the data packet, and extracts the inner IP address. The CGW then forwards the data packet to the Internet host using standard routing protocols.

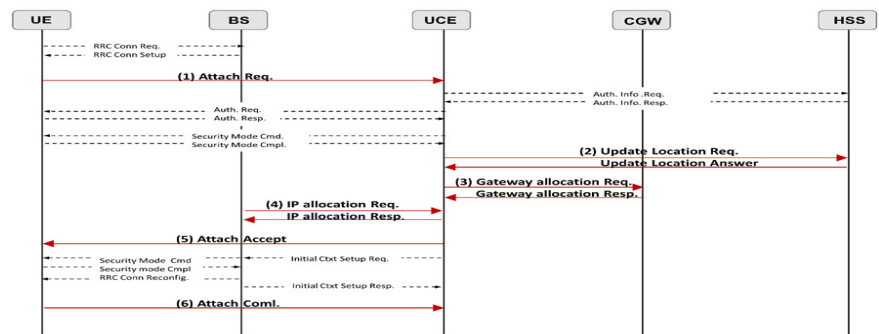

Figure 13. The initial attach procedure.

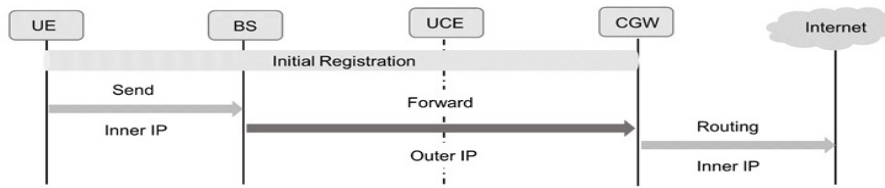

Figure 14. The data delivery procedure for data transfer from a mobile host to Internet host.

### 5.5.2. Mobile Host to Mobile Host

Figure (15) shows the data delivery procedures for data transfer between mobile hosts within the same mobile network. First, $UE_1$ sends a data packet to $BS_1$ with the destination IP address. $BS_1$ then queries the IP address of $BS_2$ from the UCE, which replies with the address of $BS_2$. $BS_1$ then encapsulates the data packet with the IP address of $BS_2$. When $BS_2$ receives the data packet, it de-encapsulates the packet to extract the inner IP, and the packet is delivered to $UE_2$.

### 5.5.3. Internet Host to Mobile Host

Figure (16) shows the data delivery procedure for data transfer from an Internet host to a mobile host. When the Internet host sends a data packet to a UE, the CGW receives the data packet using standard routing protocols. The CGW then queries UCE to determine the IP address of the BS that destination UE is registered on, and the UCE replies with the IP address of the BS. The CGW then encapsulates the data packet using this IP address (as the outer IP address). The data packet is then forwarded to the destination BS, which de-encapsulates the packet, and delivers it to the UE.

### 5.6 Handover

In ICNA a mobile network based on IP-in-IP, handover scenarios are classified as either inter-gateway handover or intra-gateway handover.

### 5.6.1. Inter-gateway Handover

Inter-gateway handover occurs between a source BS and a target BS when the UE moves but remains within the same domain. With this operation, there is the risk of data packet loss. When the UE moves to the target BS, the source BS still receives data packets from the CGW. To bridge this gap, these two BSs communicate with each other directly via GRE tunneling protocol [14]. This supports handover and prevents data packet loss during the handover process.

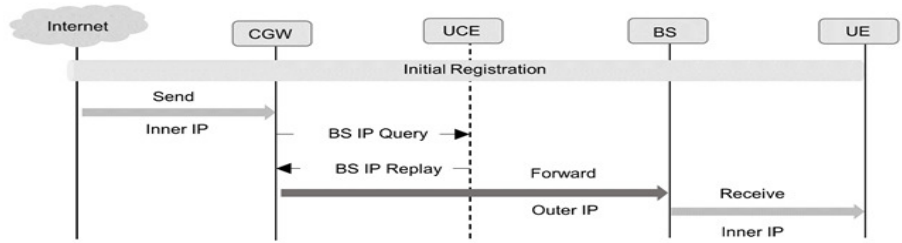

Figure 15. The data delivery procedure for data transfer from a mobile host to another mobile host.

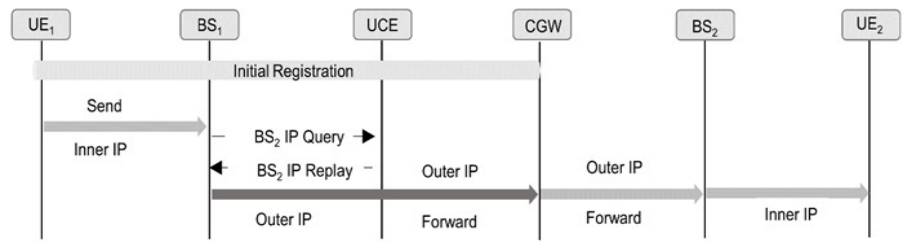

Figure 16. The data packet delivery procedure for data transfer from an Internet host to a mobile host.

Fig. 17 shows how the source BS establishes a GRE tunnel with the target BS to exchange data. During this process, the UE continues to receive data packets from the source BS until handover to the target BS has been completed.

Fig. 18 shows the protocol stack for BSs connected via a GRE tunnel. A unique GRE tunnel is generated for each UE, and each tunnel is identified by a unique key, this key is shared between the BSs.

Fig. 19 shows the handover procedure, during which the UE moves from the source BS to the target BS. The source BS sends a Handover Request to the target BS, which response with a Handover Acknowledgment, thereby establishing a GRE tunnel (step 1). The target BS sends a Path Switch Request to the UCE (step 2.a), and the UCE sends a Path Switch Response to the target BS (step 2.b). The UCE then sends a Path Modify Request to the CGW to inform the CGW that the UE has moved to a new BS, and the CGW sends a Path Modify Response to the UCE (step 3). Finally, the target BS sends a Release Resources message to the source BS, which releases the GRE tunnel.

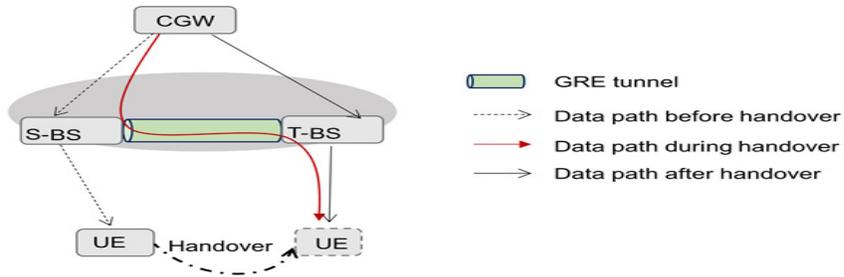

Figure 17. Establishment of a Generic Routing Encapsulation (GRE) tunnel between a source base station (BS) and a target BS.

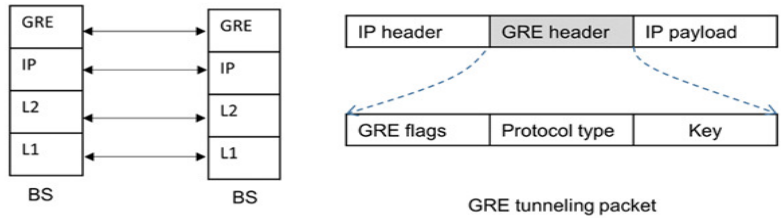

Figure 18. The GRE protocol packet stack for BSs connected via a GRE tunnel.

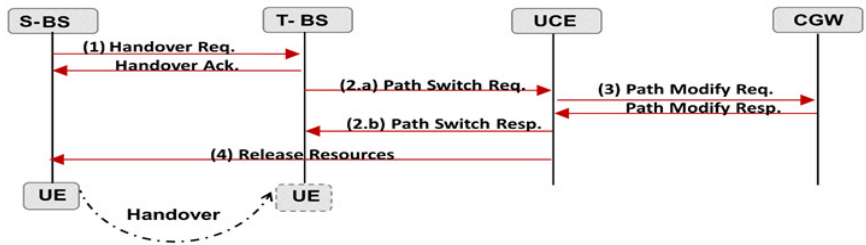

Figure 19: Inter-gateway handover process.

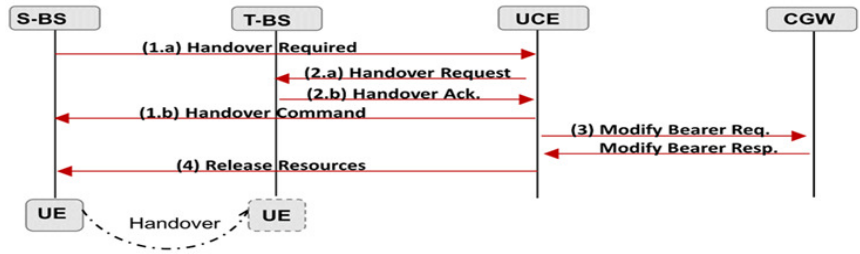

Figure 20. Intra-gateway handover process.

### 5.6.2. Intra-gateway Handover

Fig. 20 shows the intra-gateway handover process, whereby the UE moves to a different domain. During handover, the UE will move from a source BS to a target BS. The source BS will send a Handover required message to the UCE (step 1.a), which sends a Handover Request to the target BS (step 2.a). The target BS responds via a Handover Acknowledgment (step 2.b), and the UCE sends a Handover Command to the source BS (step 1.b). The UCE then sends a Modify Bearer Request to the CGW (step 3), which response with a Modify Bearer Response. Finally, the UCE sends a Release Resources message to the source BS (step 4).

## 6. NUMERICAL ANALYSIS

### 6.1. Total Transmission Delay

We calculated the total transmission delay for a centralized 4G architecture and the distributed ICNA architecture. We determined the transmission delay of a message with size S that was sent between two nodes over a wireless link and a wired link [9].

We denote the transmission delay of a message with size S sent via a wireless link from x to y as $T_{(x-y)}(S)$, which can be expressed as follows: $T_{x-y}(S) = [(1-q)/(1+q)] \times [S/B_{wl} + L_{wl}]$. We denote the transmission delay of the message with size S sent via a wired link from x node to y node as $T_{x-y}(S,H_{x-y})$, where $H_{x-y}$ represents the number of wired hops between x and y. $T_{x-y}(H_{x-y})$ can be expressed as follows: $T_{x-y}(S,H_{x-y}) = H_{x-y} \times [(S/B_w) + L_w + T_q]$. In the performance analysis, we used the notation and default parameter values listed in Table 2.

Table 2 Default parameter values.

| Parameter | Description | Value |
|---|---|---|
| $L_{wl}$ | Delay of a wireless link | 10 ms |
| $L_w$ | Delay of a wired link | 2 ms |
| q | Wireless link failure probability | 0.2 |
| $T_q$ | Average queuing delay at each node | 5 ms |
| $S_c$ | Size of control packets | 50 bytes |
| $S_d$ | Size of data packets | 200 bytes |
| $B_{wl}$ | Bandwidth of wireless links | 11 Mbps |
| Bw | Bandwidth of wired links | 100 Mbps |
| α | Hop count between eNB and SGW | 2 |
| β | Hop count between SGW and PGW | 3 |
| γ | Hop count between eNB and MME | 2 |
| δ | Hop count between MME and HSS | 3 |
| ε | Hop count between MME and SGW | 2 |
| λ | Hop count between eNBs | 2 |

### 6.1.1. 4G Network

With a 4G network, the initial attach procedure is as follows. When a UE enters an eNB region, the UE attempts to join the network by sending an Attach Request to the MME. This procedure requires an amount of time $T_{UE \rightarrow MME} = T_{UE \rightarrow eNB}(S_c) + T_{\gamma}(S_c)$. The MME registers the subscriber information and updates its location with the HSS via Update Location Request messages, and the HSS responds with an Update Location Answer. This procedure requires an amount of time $2 \times T_{\delta}(S_c)$. The MME now sends a Create Session Request to the SGW, which requires an amount of time $T_{\epsilon}(S_c)$.

The SGW establishes an EPS session with the PGW via a Modify Bearer Request, and the PGW responds with a Modify Bearer Response. This requires an amount of time $2 \times T_{\beta}(Sc)$. The SGW then responds to the MME via a Create Session Response, which requires an amount of time $T_{\epsilon}(S_c)$.

The MME performs the attach accept process with the UE by sending an Attach Accept message. This procedure requires an amount of time $T_{MME \rightarrow UE} = T_c(S_c) + T_{eNB \rightarrow UE}(S_c)$. The MME then sends an initial context message to the eNB by exchanging Initial Context Setup Request an Initial Context Response messages. This process requires an amount of time $2 \times T_{\lambda}(S_c)$. The UE then sends an Attach Complete message to the MME. This procedure requires an amount of time $T_{UE \rightarrow MME} = T_{UE \rightarrow eNB}(S_c) + T_{\gamma}(S_c)$. Finally, the MME sends a Modify Bearer Request to the SGW, which response with a Modify Bearer Response. This procedure requires an amount of time $2 \times T_{\epsilon}(Sc)$. For data delivery, data packets are sent from the UE to the eNB, which then sends the packets to the SGW. The SGW forwards the data packets to the PGW. The total transmission delay of the 4G network is given by

$$TTD_{4G} = 4\ T_{UE \rightarrow eNB}(S_c) + 5\ T_{\gamma}(S_c) + 2T_{\delta}(S_c) + 4T_{\epsilon}(S_c)$$
$$2T_{\beta}(S_c) + 2T_{UE \rightarrow eNB}(S_d) + 2T_{\epsilon}(S_d) + 2T_{\beta}(S_d) \tag{1}$$

### 6.1.2. ICNA Network

With the ICNA network, the initial attach procedure is as follows. When the UE enters a BS region, it attempts to join the network by sending an Attach Request to the UCE. This procedure requires an amount of time $T_{UE \rightarrow UCE} = T_{UE \rightarrow BS}(S_c) + T_{\gamma}(S_c)$. The UCE registers the subscriber information and updates its location with the HSS by sending an Update Location Request, and the HSS responds with an Update Location Answer. This procedure requires an amount of time $2 \times T_{\delta}(S_c)$. The UCE then sends a Gateway Allocation Request to the CGW, which response by sending a Gateway Allocation Response. This procedure requires an amount of time $2 \times T_{\epsilon}(S_c)$.

The BS requests IP address allocation from the UCE by exchanging IP Allocation Request and IP Allocation Response messages. This procedure requires an amount of time $2 \times T_{\gamma}(S_c)$. The UCE then performs the attach accept procedure with the UE, and then sends an Attach Accept message to the UE. This procedure requires an amount of time $T_{UCE \rightarrow UE} = T_{\gamma}(S_c) + T_{BS \rightarrow UE}(S_c)$. The UE now sends an Attach Complete message to UCE, which requires an amount of time $T_{UE \rightarrow UCE} = T_{UE \rightarrow BS}(S_c) + T_{\gamma}(S_c)$. For data delivery, the data packets sent from the UE to the BS are then sent to the CGW. Thus, the total transmission delay for ICNA is as follows:

$$TTD_{ICNA} = 3T_{UE \rightarrow BS}(S_c) + 5T_{\gamma}(S_c) + 2T_{\delta}(S_c) + 2T_{\epsilon}(S_c)$$
$$+ 2T_{UE \rightarrow BS}(S_\delta) + 2T_{\alpha}(S_d) \tag{2}$$

## 6.2. Handover Delay

### 6.2.1. Handover Delay in a 4G Network

a)  X2 Handover with S-GW relocation

In a 4G network, when the UE moves to another eNB region, the source eNB sends a *Handover Request* to the target eNB, which response with a *Handover Acknowledgment*. The target eNB then sends a *Path Switch Request* to the MME, which sends a *Create Session Request* to the SGW. The SGW exchanges modify bearer information with the P-GW via *Modify Bearer Request* and *Modify Bearer Response* messages. The SGW then sends a *Create Session Response* to the MME, which sends a *Path Switch Response* to the target eNB. Finally, the MME sends a *Release Resources* message to source the eNB. The total X2 handover delay of the 4G network is as follows:

$$XHD_{4G} = 2T_{\lambda +} 3T_{\gamma} + 2\ T_{\epsilon} + 2T_{\beta} \tag{3}$$

b)  S1 handover with S-GW relocation

With S1 handover, when the UE moves to a different domain, the source eNB sends a *Handover required* message to the MME, which in turn sends a *Handover Request* to the target eNB. The target eNB respond with a *Handover Acknowledgment*, and then the MME sends a *Handover Command* to the source eNB. The target eNB sends a *Handover Notify* message to MME, which sends a *Modify Bearer Request* to the target SGW. The SGW exchanges modify bearer messages with the PGW via *Modify Bearer Request* and *Modify Bearer Response*. The target SGW then sends a *Modify Bearer Response* to MME, which sends a *Release Resources* request to the source eNB. The S1 handover delay of 4G ($SHD_{4G}$) is as follows:

$$SHD_{4G} = 6T_{\lambda} + 2T_{\epsilon} + 2T_{\beta} \tag{4}$$

### 6.2.2. Handover in ICNA Network

a)  Inter-gateway handover

With inter-gateway handover, when the UE moves to another BS region, the source BS sends a *Handover Request* to the target BS, which response with a *Handover Acknowledgment*. The target BS sends a *Path Switch Request* to the UCE, which sends a *Path Modify Request* to the CGW. The CGW sends a *Path Modify Response* to the UCE, which sends a *Path Switch Response* to the target BS. Finally, the target BS sends a *Release Resources* message to the source BS. We obtain the inter-gateway handover delay of ICNA network as follows:

$$IGHD_{ICNA} = 3T_{\lambda} + 2T_{\gamma} + 2T_{\epsilon} \tag{5}$$

b) Intra-gateway handover

With intra-gateway handover, when the UE moves to a different domain, the source BS sends a *Handover required* message to the UCE, which sends a *Handover Request* to the target BS. The target BS responds by sending a *Handover Acknowledgment* and the UCE then sends a *Handover Command* to the source BS. The UCE also sends a *Modify Bearer Request* to the CGW, which response via a *Modify Bearer Response*. Finally, the UCE sends a *Release Resources* request to the source BS. We obtain the intra-gateway handover delay of an ICNA network as follows:

$$IAGHD_{ICNA} = 5T_{\gamma} + 2T_{\epsilon} \tag{6}$$

## 6.3. Data Tunneling Overhead

With GTP used in 4G mobile networks, data packets are encapsulated with three headers: an 8-byte GTP header, an 8-byte UDP header, and a 20-byte IP header, which totals 36 bytes [15]. In an ICNA network, the data packets are encapsulated with a 12-byte outer IP header and a 20-byte inner IP header, which totals 32 bytes. Furthermore, with an ICNA network, GRE tunneling protocol is used for handover between eNBs, whereby data packets are encapsulated with an 8-byte GRE header and a 20-byte IP header, which totals 28 bytes. We may therefore obtain the data tunneling overheads of a 4G network ($DTO_{4G}$), an ICNA network ($DTO_{ICNA}$), and a GRE protocol ($DTO_{GRE}$) as follows:

$$DTO_{4G} = \frac{GTP/UDP/IP}{GTP/UDP/IP + Data\ packet\,(Sd)} \times 100$$

$$DTO_{ICNA} = \frac{Outer\ IP/Inner\ IP}{Outer\ IP/Inner\ IP + Data\ packet\,(Sd)} \times 100$$

$$DTO_{GRE} = \frac{GRE/IP}{GRE/IP + Data\ packet\,(Sd)} \times 100$$

### 6.4. Numerical Results

Based on the mathematical relationships given above, we may compare the performance of the existing 4G network architecture with that of the proposed ICNA architecture. Table 2 lists the default values of the delay parameters, which were taken from Ref. [16].

### 6.4.1. Total transmission delay

Figure (21) shows the total transmission delay as a function of the average queuing delay at each network node, $T_q$. The total transmission delay increased linearly as $T_q$ increased for both networks. This shows that the 4G network results in worse performance than the ICNA network. This is because the PGW implements data packet delivery to the SGW in the 4G network, whereas data packets are delivered via the optimal path with the ICNA network.

Figure (22) shows the total transmission delay as a function of the wireless link delay, $L_{wl}$. For both networks, the total transmission delay increased linearly as $L_{wl}$ increased. Again, the ICNA network resulted in superior performance to that of the 4G network. Figure (23) shows the effects of the hop count between the eNB and MME. Varying $\gamma$ had a significant effect on the transmission delay of the 4G network. This is because data packets are routed via a centralized anchor, such as SGW and MME. However, with the ICNA network, the total delay was not significantly affected by $\gamma$, since data packets were delivered between the CGW and the BS via an optimized route.

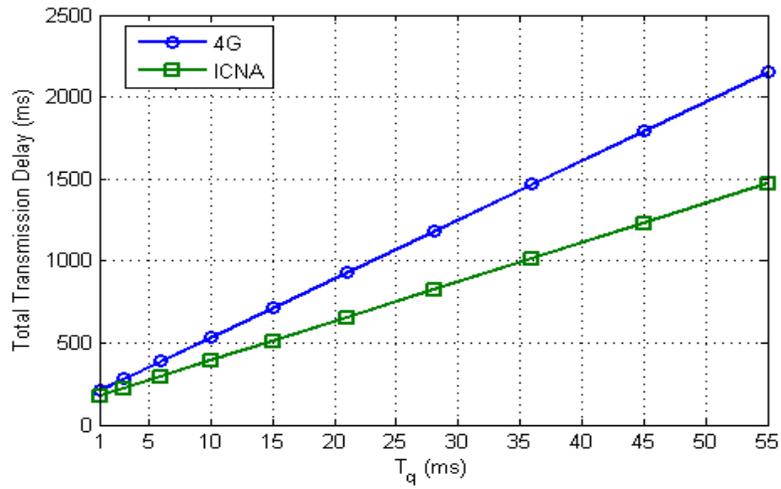

Figure 21. The effect of the queuing delay at each node.

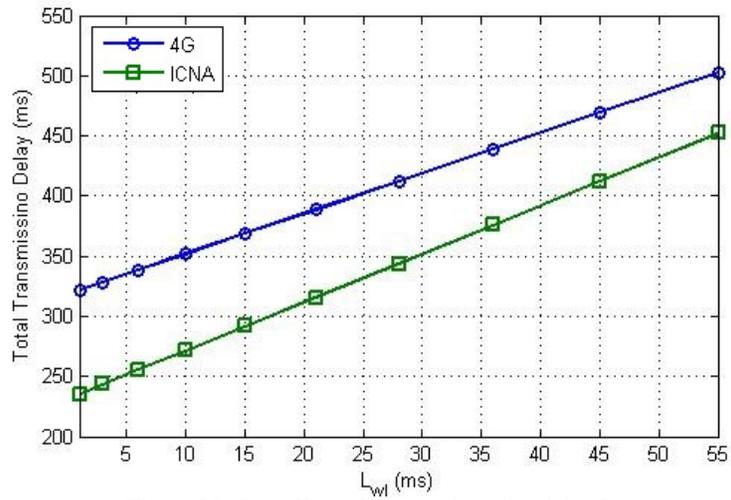

Figure 22. The effect of the wireless link delay $L_{wl}$.

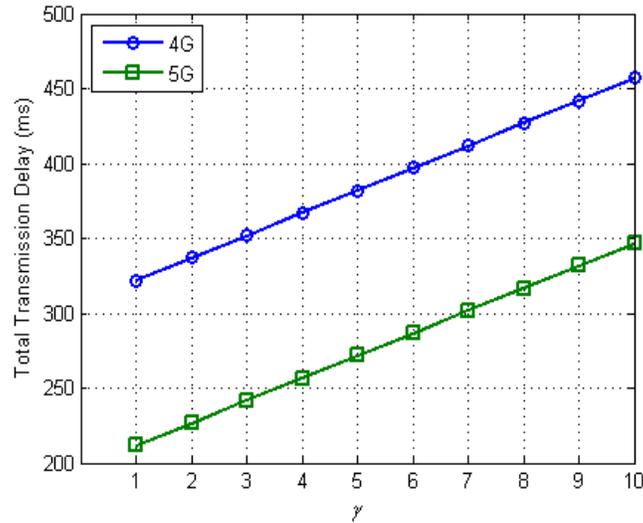

Figure 23. The effect of varying γ on the total transmission delay.

### 6.4.2. Handover delay

Figure (24) shows a comparison of inter-gateway handover between an ICNA and X2 handover with S-GW relocation in a 4G network. This comparison is based on varying the hop count $\lambda$ between eNBs in the same domain. Varying $\lambda$ had a greater impact on the handover delay for the 4G network. This is because the MME modifies the bearer via a centralized anchor (SGW/PGW), whereas the ICNA is largely insensitive to $\lambda$ because the UCE performs path modification with the CGW.

Figure (25) shows a comparison of intra-gateway handover between an ICNA with S1 handover with S-GW relocation in a 4G network. This comparison was based on the hop count γ between the eNB and MME. Varying c had a significant impact on the handover delay for the 4G network; this is because the MME implements modify bearer with the PGW via the SGW. By contrast, with the ICNA network, the UCE performs path switching directly with CGW.

### 6.4.3. Data Tunnelling Overhead

Figure (26) and (27) show comparisons between 4G and ICNA networks in terms of the data tunneling overhead. Figure (26) shows the data tunneling overhead for various payload sizes for the 4G GTP protocol and the ICNA protocol. The payload size significantly affected the data tunneling overhead for both network protocols. This is because the GTP protocol encapsulates the payload using three headers (GTP/UDP/IP), totaling 36 bytes, and the ICNA network protocol encapsulates the payload using two headers (outer IP and inner IP), totaling 32 bytes.

Figure (27) shows a comparison of GTP with the GRE protocol, which is used to support handover between BSs. The payload significantly affected the data tunneling overhead for both protocols. This is because the headers that are added to the payload total 28 bytes with the GRE protocol and 36 bytes with GTP.

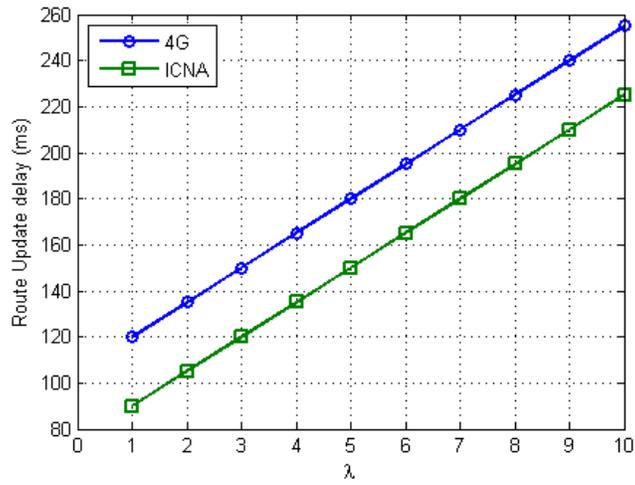

Figure 24. The effect of λ on the handover delay.

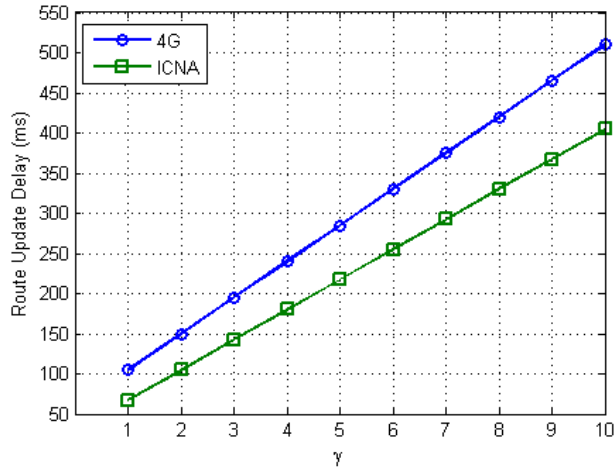

Figure 25. The effect of γ on the handover delay.

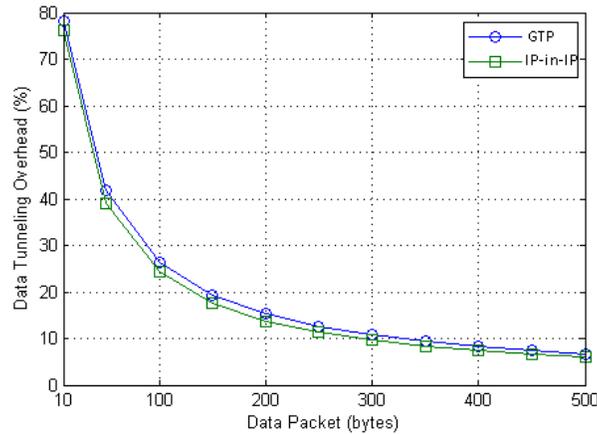

Figure 26. The effect of Sd on the data tunneling overhead (General Packet Radio Service tunneling protocol [GTP] vs. IP-in-IP).

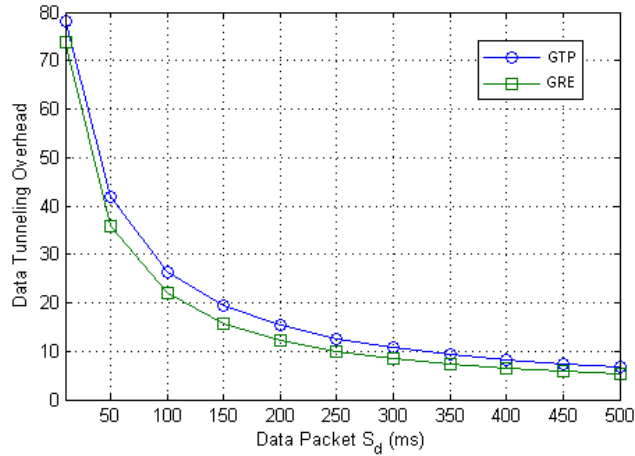

Figure **27.** The effect of Sd on the data tunneling overhead (GTP vs. GRE).

# 7. SIMULATION RESULTS

We evaluate the performance of the proposed model by using NS-3 simulation. For 4G network many simulations have been proposed i.e. LTE/EPC Network Simulator (LENA) project [17]. We rebuild the current LENA simulation as described in [18]. Besides, we used the implementation previously described to run an ICNA simulation. NS-3 version 3.22 in Linux environment has been used. The programming of proposed ICNA consists of building BS, UCE, CGW and L3S. Control plane and data plane implementing into the corresponding interface. IP-in-IP protocol implemented in the core network. A remote host node is created to acts as an Internet server, which is able to send packets to other nodes. The rest of the simulation parameters are configured as shown in the table (3).

Table 3 simulation key values

| Parameter | Setting |
|---|---|
| Number of UE | One node |
| Speed of UE | Varies from 5to 120 km/h |
| eNB Tx power | 46dBm |
| Distance between eNB | 100 meters |
| EPS Bearer | NGBR-VIDEO-TCP |
| QoS Class Identifier (QCI) | 9 |
| Bandwidth | 5MHz |
| Data rate | 10Gbps |

This simulation builds to measure the performance between the 4G network and the proposed schema as a following: time for data delivery between the mobile host and Internet host, time for data delivery between a mobile host in the same network and Intra-gateway handover.

Figure (28) shows the total latency of initial attachment and data delivery from the mobile host to an Internet host. The total transmission delay increased linearly for both networks. This shows that the 4G network results in worse performance than the ICNA network. This is because the PGW implements data packet delivery to the SGW in the 4G network, whereas data packets are delivered via the optimal path with the ICNA network.

Figure (29) shows a comparison of the handover delay between an ICNA and X2 handover with S-GW relocation in a 4G network. This comparison is based on the number of eNBs in the same domain. The simulation shows a greater impact on the handover delay for the 4G network. This is because the MME modifies the bearer via a centralized anchor (SGW/PGW), whereas the ICNA is largely insensitive to handover delay because the UCE performs path modification with the CGW.

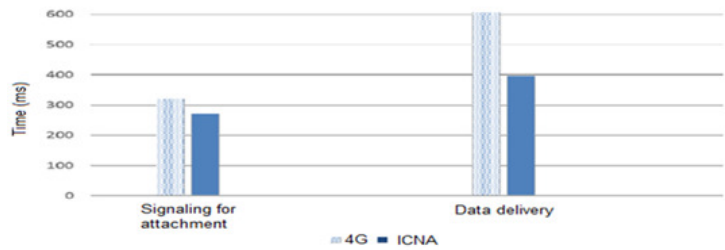

Figure **28**. Comparison of 4G and proposed ICNA in term of latency of initial attachment and data delivery from mobile host to Internet

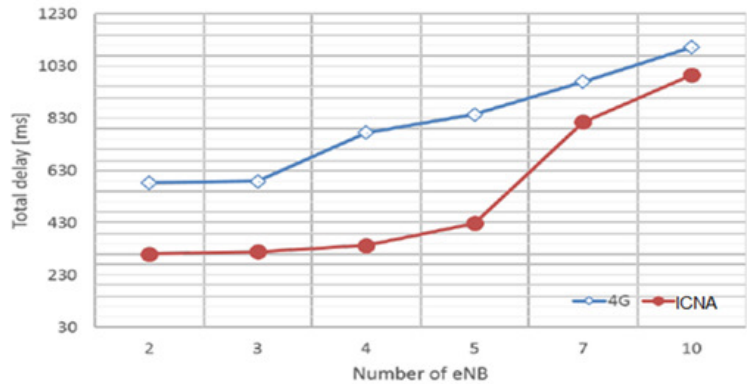

Figure **29.** Comparison of 4G and proposed ICNA in term of latency of Handover

## 8. CONCLUSIONS

The current 4G mobile packet core (i.e., EPC) is inflexible and expensive. We have described an ICNA network architecture, which is based on IP-in-IP protocols for data packet delivery. With this architecture, the mobile packet core nodes function as layer 3 switches, and data packets are delivered between eNBs and PDN gateways via routing protocols.

ICNA achieves better performance than 4G in three ways. First, it eliminates the GTP tunneling protocol and using IP-in-IP protocols. Second, there is no centralized mobility anchor because the data traffic is distributed on the network core. Third, the ICNA protocol is more efficient than 4G because the core network is based on L3S.

We compared our ICNA architecture with the existing 4G network architecture using numerical analyses and simulation. The results show that the ICNA architecture results in better performance than the 4G network in terms of the total transmission delay, the handover delay and the data tunneling overhead.

## AUTHORS

**Mohammad Al Shinwan,** received his B.S degree in Computer Science from Al al-Bayt
university, Jordan, in 2009 and the master degree in 2013 from the institute of Mathematical
and Computer Science, University of Sindh in Pakistan. He received a Ph.D. in Computer
Networks from Inje University, Korea. He is currently an assistant professor in the Faculty
of Computer Science and Informatics, Amman Arab University, Jordan. His current research
interests include mobility management, network management and OAM (Operation,
Administration and Maintenance) for future mobile network.

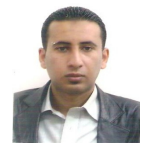

**Chul-Soo Kim** is a professor in the School of Computer Engineering of Inje University in
Gimhae, Korea. He received Ph.D. from the Pusan National University (Pusan, Korea) and
worked for ETRI (Electronics and Telecommunication research Institute) from 1985 - 2000
as senior researcher for developing TDX exchange. Aside from the involvement in various
national and international projects, his primary research interests include network protocols,
traffic management, OAM issue, and NGN charging. He is a member of ITU-T SG3, SG11,
SG13 and a Rapporteur of ATM Lite from 1998 - 2002, and CEO in WIZNET from 2000 - 2001. He is
currently the chairperson of BcN Reference Model in Korea, and a Rapporteur of ITU-T SG3 NGN
Charging.

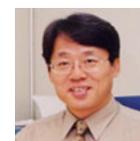